\documentclass{article}

\usepackage{arxiv}

\usepackage[utf8]{inputenc} 
\usepackage[T1]{fontenc}    
\usepackage{hyperref}       
\usepackage{url}            
\usepackage{booktabs} 
\usepackage{multirow}
\usepackage{amsfonts}       
\usepackage{amsmath}
\usepackage{nicefrac}       
\usepackage{microtype}      
\usepackage{cleveref}       
\usepackage{lipsum}         
\usepackage{graphicx}
\usepackage{doi}
\usepackage{color}

\usepackage{algorithm}
\usepackage{algpseudocode}
\usepackage{dsfont}
\usepackage{subcaption}

\newcommand{\calR}{\mathcal{R}}
\newcommand{\calD}{\mathcal{D}}
\newcommand{\bbR}{\mathbb{R}}

\title{RARe: Raising Ad Revenue Framework with Context-Aware Reranking}

\date{}

\DeclareMathOperator*{\argmax}{argmax}

\newif\ifuniqueAffiliation

\ifuniqueAffiliation 
\author{ 
}
\else
\usepackage{authblk}

\newbox{\orcid}\sbox{\orcid}{\includegraphics[scale=0.06]{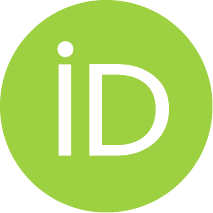}} 
\author[2, 1]{%
	\href{https://orcid.org/0009-0006-2279-5926}{\usebox{\orcid}\hspace{1mm}Ekaterina Solodneva
    }%
}
\author[1, 3]{%
	\href{https://orcid.org/0000-0002-2142-3008}{\usebox{\orcid}\hspace{1mm}Alexandra Khirianova
    }%
}
\author[5]{%
	\href{https://orcid.org/0000-0002-7050-1615}{\usebox{\orcid}\hspace{1mm}Aleksandr Katrutsa
    }%
}
\author[1]{%
	\href{https://orcid.org/0009-0001-1700-7204}{\usebox{\orcid}\hspace{1mm}Roman Loginov
}%
}
\author[1]{%
	\href{https://orcid.org/0009-0001-1776-0116}{\usebox{\orcid}\hspace{1mm}Andrey Tikhanov
}%
}
\author[1]{%
	\href{https://orcid.org/0000-0003-3399-4235}{\usebox{\orcid}\hspace{1mm} Egor Samosvat
}%
}
\author[4, 2]{%
	\href{https://orcid.org/0000-0003-0533-3018}{\usebox{\orcid}\hspace{1mm} Yuriy Dorn
}%
}
\affil[1]{Avito, Moscow, Russia}
\affil[2]{Moscow Institute of Physics and Technology, Moscow, Russia}
\affil[3]{Lebedev Physical Institute of the Russian Academy of Sciences, Moscow, Russia}
\affil[4]{Lomonosov Moscow State University, Moscow, Russia}
\affil[5]{AIRI, Moscow, Russia}
\fi

\begin{document}
\maketitle

\begin{abstract}
Modern recommender systems excel at optimizing search result relevance for e-commerce platforms. 
While maintaining this relevance, platforms seek opportunities to maximize revenue through search result adjustments. 
To address the trade-off between relevance and revenue, we propose the \textsf{RARe} (\textbf{R}aising \textbf{A}dvertisement \textbf{Re}venue) framework.
\textsf{RARe} stacks a click model and a reranking model. 
We train the \textsf{RARe} framework with a loss function to find revenue and relevance trade-offs.
According to our experience, the click model is crucial in the \textsf{RARe} framework.
We propose and compare two different click models that take into account the context of items in a search result. 
The first click model is a Gradient-Boosting Decision Tree with Concatenation (GBDT-C), which includes a context in the traditional GBDT model for click prediction.
The second model, SAINT-Q, adapts the Sequential Attention model to capture influences between search results.
Our experiments indicate that the proposed click models outperform baselines and improve the overall quality of our framework.
Experiments on the industrial dataset, which will be released publicly, show \textsf{RARe}'s significant revenue improvements while preserving a high relevance.
\end{abstract}

\section{Introduction}
Revenue optimization in e-commerce requires careful balancing between user experience and monetization strategies. 
E-commerce platforms generate revenue through various mechanisms, including sales commissions, promoted products, and advertising placements. 
While revenue maximization is fundamental to platform operations, recommender systems are typically optimized for relevance metrics only. 
This effectiveness in user engagement creates an opportunity to achieve higher revenue by carefully adjusting recommendation outputs. 
This work focuses on revenue maximization through promoted products placement. 
The key challenge is to determine optimal positions for promoted items. 
To address the trade-off between relevance and revenue, we propose the \textsf{RARe} framework.

We focus on user engagement metrics, specifically click-through rate (CTR), as a direct indicator of platform revenue generation.
As our goal is to maximize CTR on promoted items, it is crucial to understand the mechanics of user interactions.
While users' behavior provides the most accurate insights into CTR drivers, collecting comprehensive and reliable user data is often difficult.
Therefore, we construct a click model based on methodologies from studies~\cite{monster, 2021context, 2020relevance}.
In this work, we define a click model as a machine learning model that simulates user behavior by predicting click probabilities on search results.

Traditional search result optimization assumes independent item evaluation.
This assumption leads to the widely known pointwise approach~\cite{pointwise}.
However, real user behavior shows that neighboring items significantly affect user behavior~\cite{effects,banner,monster}. 
Therefore, incorporating a listwise approach within the search result page into the click models becomes essential for emulating user behavior, see~\cite{2021context,2020relevance}.
Furthermore, recent models~\cite{lstm_with_attn_CTR,xu2021disentangled} have shown promising results in CTR prediction by incorporating user context, including platform, time, location, and item position. 
We also use contextual features to capture the dependencies between items and user behavior.
Hence, by a \textbf{context-aware} model, we mean both the consideration of \emph{contextual features} and the adoption of a \emph{listwise} approach.

After selecting the click model, the next challenge is determining the optimal placement for promoted items.
The most straightforward approach involves assessing all possible permutations of item placements.
However, this approach is computationally infeasible due to the large number of possible positions in search results.
Recent deep learning methods, which have been widely used in recommendation systems recently~\cite{madanchian2024impact}, could provide a solution to this problem.
Previous study~\cite{2021context} has already demonstrated the successful use of MLP as a reranking model that generates locally optimal placement for promoted items.

Balancing revenue and user experience remains a central challenge for e-commerce platforms. 
The \textsf{RARe} framework proposes a context-aware solution with significant potential to address this trade-off. 
By optimizing promoted product placement using contextual information, \textsf{RARe} offers a promising approach for enhanced revenue generation without a drop in user engagement. 
The following sections will explore the architecture of our framework and empirically confirm its potential to impact e-commerce monetization.

Our main contributions are summarized as follows:

\begin{itemize}
\item We propose a novel data processing approach for the Gradient Boosting Decision Tree algorithm that makes it context-aware for the CTR prediction problem. 

\item We develop the context-aware modification of the transformer-based SAINT model for tabular data that makes it appropriate for the CTR prediction problem.

\item  We design the \textsf{RARe} framework for context-aware revenue maximization, where the reranking model generates adjustment of the search results order.
The reranking model is trained with the developed clicker models.

\item We release a public dataset for benchmarking revenue optimization and CTR prediction tasks, fostering further research in this domain.
\end{itemize}

\section{Related works}
Typically, revenue maximization is solved using standalone reranking models. 
For example, studies~\cite{kompan2021exploring, de2024model,pei2019value} propose models based on individual-based metrics that improve revenue but ignore the influence of neighboring items. 
The latter feature can lead to a reduction in user engagement if high-value items are placed together at the top of the page with search results.
Therefore, standalone reranking models may permute search results in a way that increases revenue but decreases user engagement.
To prevent this negative effect, an auxiliary click model should be used to evaluate the relevance of the perturbed search results generated by the re-ranking model~\cite{2021context}.
The standard quantity for relevance estimation is click-through rate (CTR).
The industry standard for CTR prediction~\cite{chen2024branches, gao2024gbdt4ctrvis} is
Gradient Boosting Decision Trees (GBDT) ~\cite{friedman2001greedy,chen2016xgboost,prokhorenkova2018catboost,ke2017lightgbm}.
At the same time, factorization machines, like DeepFM~\cite{guo2017deepfm} and FFM~\cite{juan2016field}, improve CTR prediction accuracy by explicitly capturing the dependence between items' features.
In contrast, deep learning models like Wide \& Deep~\cite{cheng2016wide} and FiBiNET~\cite{huang2019fibinet} 
rely on neural networks to implicitly capture dependence between items' features through non-linear transformations.
Although Transformer models~\cite{stec, hou, rat, chen2022extr} could fit the CTR prediction task, they still track only dependencies within item features. 
Thus, the approaches mentioned above do not consider the user's features like previous clicks, timestamp, location, etc.

To address this limitation, models for processing sequential users' actions are developed.
In particular, Recurrent Neural Networks (RNN)~\cite{lstm_with_attn_CTR} and attention mechanism~\cite{vaswani2017attention,min2022neighbour} improve CTR prediction in this setup. 
Other RNN-based models like DIN~\cite{din} and DIEN~\cite{dien} predict user actions based on the available history logs.
While the mentioned studies include user features and the corresponding sequential data in the models for CTR prediction, they do not consider how the neighbor items affect the CTR of the selected item. 
This effect is important for users' behavior, see~\cite{effects,banner}.

A combination of the clicker and reranking models is proposed in~\cite{monster}, where the Bi-GRU clicker model captures the search context and the RL-based reranking model with GRU and MLP adjusts the search results order. 
However, this model training is too costly due to the architecture's complexity, and the authors do not share their implementation. 
A similar idea is developed in~\cite{2021context}, where Position-aware Graph Embedding with Bi-LSTM is used for modeling item interactions while contextual loss is used for pairwise revenue maximization. 
This approach effectively captures context over the search results items but suffers from processing long sequences of items in search results. 

\section{Problem Statement}
\label{sec::problem_statement}
Let $I_q = \{i_1, \ldots, i_{N}\}, i_j \in \mathbb{N}$ be the ordered set of items ids generated by the pre-trained model based on the user query~$q$.
By construction, the item ids are ordered in descending order with respect to their relevance to the corresponding query~$q$.
The pre-trained model maximizes the relevance of the entire set $I_q$.
However, this purpose may contradict the platform revenue maximization. 
Thus, initially, any platform could solve the bi-objective optimization problem, where the objective functions are the relevance of the search results and the revenue that these results could provide.

At the same time, relevance is the base metric for any platform since it corresponds to user engagement.
Therefore, instead of the multi-objective optimization framework~\cite{lin2019pareto,katrutsa2020follow}, we use the optimal perturbation search paradigm.
In this paradigm, we need to perturb the order of elements from $I_q$ such that the revenue becomes larger while relevance drops slightly.

Formally, denote by $\pi_0$ the permutation of the items ids corresponding to the order in the set $I_q$.
Then the revenue maximization problem reads as follows
\begin{equation}
    \begin{split}
        &\pi^* = \argmax_{\pi} \calR(\pi)\\
        \text{subject to } & \calD(\pi, \pi_0) \leq \tau, \\ 
    \end{split}
    \label{eq::general_problem}
\end{equation}
where $\calR$ denotes the revenue generated by the permutation $\pi$ and $D$ denotes the relevance difference between two permutations and $\tau > 0$ is a pre-defined threshold.
To simplify the search of the permutation $\pi$, we introduce the reranking model $R$ parametrized by the vector $\theta \in \bbR^n$, where $n$ is a number of trained parameters.
This model generates the permutation of items from $I_q$ with given features of query $q$, initial permutation $\pi_0$, and features of items from $I_q$. 
Then, problem~(\ref{eq::general_problem}) is reduced to the following one
\begin{equation}
    \begin{split}
        &\theta^* = \argmax_{\theta} \calR(\theta)\\
        \text{subject to } & \calD(R(\theta), \pi_0) \leq \tau. \\ 
    \end{split}
    \label{eq::parametrized_problem}
\end{equation}
To train the reranking model, one needs a set of queries $Q = \{q_1, \ldots, q_M\}$, the base search results $S = \{ I_{q_1}, \ldots, I_{q_M}\}$ for them and the revenues generated by clicking on every item $\mathbf{r} \in \mathbb{R}^{C}_+$, where $C$ is a total number of available items.
In addition, the $i$-th item is described via the corresponding vector of features $\mathbf{e}_i \in \bbR^d$, where $d$ is a number of item features or the dimension of the embedding.
Now, we can formally define how to compute revenue $\mathcal{R}$ and difference $\calD$.

\paragraph{Revenue.} 
To compute the revenue generated by the given search results $I_q$, one has to know what items are clicked by users since only these items give revenue to the e-commerce platform.
Since we do not know what items will be clicked in advance, we estimate the probability $p_i$ of click to the $i$-th item from $I_q$ and then use these probabilities in the revenue estimate.
To estimate click probabilities, we introduce the \emph{clicker model} that is trained based on the available users' action logs consisting of queries, search results, and labels $\{ 1, 0\}$ corresponding to clicked and not clicked items. 
Based on $p_i$ estimated by the clicker model and pre-defined threshold $h > 0$, we identify the items to be clicked with the vector $\mathbf{c} \in \{0, 1 \}^N$ such that 
$c_i = \begin{cases}
1, & \text{$p_i$} \geq h\\
    0, & \text{$p_i$} < h.
\end{cases}$
Now, we have estimated the items that will be clicked with high probability and know the revenue from clicking on every item: $r_i$.
Therefore, the total revenue can be estimated as
\[
\calR(\theta) = \frac{1}{|Q|} \sum_{q \in Q} \sum_{i \in \widehat{I}_q} r_i c_i,
\]
where a reranking model $R$ with parameters $\theta$ generates search results $\widehat{I}_q$.

\paragraph{Relevance.}
The next key ingredient of our problem statement is the difference between the perturbed order of items from search results and the original order.
To estimate the impact of such difference on user experience, study~\cite{2020relevance} suggests using KL-divergence between the corresponding normalized relevance scores for items in two permutations.
Since we assume that the original search result $I_q$ is already sorted from the most relevant items to the least relevant ones, we define the relevance $v$ as a monotonically decreasing function of the item's position $j$.
We use the following definition of relevance: $v_j = P^{j-1}$, where $P\in[0,1]$ is a coefficient that simultaneously captures two effects: (1) the decrease in item relevance with position and (2) the lower position of the item is, the smaller probability that user clicks on it. 
This leads to the following formula for calculating the Difference metric:
\begin{equation}
    \calD(\pi, \pi') = \frac{1}{|Q|} \sum_{q \in Q} \sum_{j=1}^{|I_q|} \hat{v}_j \log \frac{\hat{v}_j}{v_j},
    \label{eq::kl_difference}
\end{equation}
where $\hat{v}_i$ is the original vector $v$, reordered by the reranking model, according to the permutation $\pi'$.
Permutations of items from the bottom have a smaller impact on the metric $\calD$ compared to permutations of items from the top positions. 
This feature is important since users typically scan the items from the top positions of the search results.


\paragraph{Reranking model training.}
According to the previous discussion, the reranking model generates perturbation of the original order of items.
To simplify the training of the reranking model, we replace the explicit perturbation generation with the scores $s_j$, which the reranking model assigns to every item from $I_q$.
The higher the score is, the higher the position of the item in the perturbed order is.
We use the pre-trained and frozen clicker model to train a reranking model and follow the pipeline suggested in~\cite{2021context}.
This pipeline consists of two important ingredients.
The first ingredient is the loss function
\begin{equation}
    \begin{split}
        & \mathcal{L}_\lambda(\pi, \pi') = \sum_{j=1}^N \sum_{j' = 1}^N  \mathds{1}[i_j > i_{j'}]\Delta \calR_{abs}(j', j) \log(1 + e^{-\sigma(s_j - s_{j'})}),\\
        & \Delta \calR_{abs} (j', j) = \calR_{abs} (\pi', j') - \calR_{abs} (\pi, j), 
    \end{split}
    \label{eq::loss}
\end{equation}
where $\sigma$ denotes a sigmoid function, $\mathds{1}$ denotes the indicator function, $\pi$ is a search result order based on the scores $s_j$ and $\pi'$ is the same as $\pi$ up to the single transposition of items $i_j$ and $i_{j'}$.
The larger the difference between the scores $s_j$ and $s_{j'}$, the far the items from the reranking model order. 
Therefore, each transposition with a high $\Delta\calR_{abs}$ and a large score difference contributes significantly to the loss. 
Thus, the loss function penalizes such reranking model permutations $\pi$ that could be modified with a single transposition of the top items, and the modified order of items provides larger revenue.

The second ingredient follows from the loss function definition and includes the procedure of generating trial neighbor permutations for $\pi$.
Every trial permutation differs only by a single transposition.
The set of all such trial permutations is crucial for searching the locally optimal parameters $\theta^*$ of the reranking model through the loss function~(\ref{eq::loss}). 

Note that \textsf{RARe} does not directly incorporate the constraint on Difference (see~(\ref{eq::parametrized_problem})) in the training of the reranking model. 
However, our framework is flexible enough to adjust the Difference through an additional parameter, see Section~\ref{sec::reranker}.

\section{\textsf{RARe} framework architecture}
As previously mentioned, our \textsf{RARe} framework consists of the clicker and reranking models.
The key feature of our framework is tracking the influence of the neighboring items on the reranking model output.
Since the Gradient Boosting Decision Trees (GBDT) and Transfomer-based models are the most promising for click prediction, we consider below their context-aware modifications, where we mean by context the neighbors of the items in the search result. 

\subsection{GBDT-C: context-aware GBDT clicker model}
\label{sec::clickers}
\begin{figure}[!h]
    \centering
    \begin{subfigure}[b]{0.2\textwidth}
        \centering
\includegraphics[width=\textwidth]{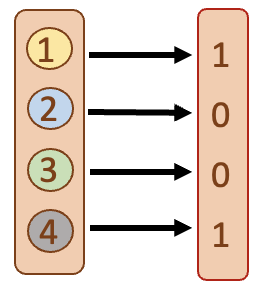}
        \caption{GBDT}
        \label{fig::gbdt_scheme_a}
    \end{subfigure}
    \qquad \qquad
    \begin{subfigure}[b]{0.45\textwidth}
        \centering
        \includegraphics[width=\textwidth]{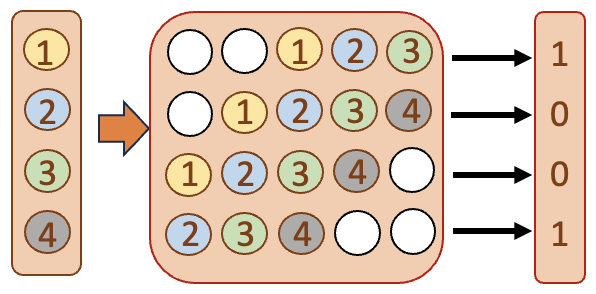}
        \caption{GBDT-C}
        \label{fig::gbdt_scheme_b}
    \end{subfigure}
    \caption{Comparison of the classical GBDT model (left) and the proposed GBDT-C model (right). 
    The GBDT-C model requires expanding the original dataset's feature dimension by stacking the neighbor items' features.
    The colored circles represent the features of the items, the black arrows correspond to the model click model processing of the input features, and $1/0$ on the right indicate the model predictions.}
\label{fig::gbdt_scheme}
\end{figure}

To construct the context-aware GBDT model, we consider the CatBoost~\cite{prokhorenkova2018catboost} implementation of the GBDT algorithm as a basis and compare our modification with it.
The natural drawback of the classical GBDT algorithm inherited in the CatBoost implementation is that it does not consider the features of the adjacent items within the search results. 
Therefore, it predicts clicks solely based on the features of the individual items. 
Our context-aware GBDT clicker model (GBDT-C) differs from the basic algorithm by the specific train dataset preparation.
In particular, we propose to enrich items features with features from items whose positions are $k$ slots above and $k$ slots below the target item, thereby enhancing the accuracy of click predictions (see Figure ~\ref{fig::gbdt_scheme}).
More formally, for each item $i_j \in I_q, j=1, \ldots, N$, features from the $k$ preceding and $k$ following neighbors are stacked as $\widehat{\mathbf{e}}_j = [\mathbf{e}_{j-k}, \ldots \mathbf{e}_{j-1}, \mathbf{e}_j, \mathbf{e}_{j+1}, \ldots, \mathbf{e}_{j+k}]$.
The modified feature representations of the items are then stacked into a dataset, where each item $i_j$ is described via combined feature set $\widehat{\mathbf{e}}_j$.
The dependence of the performance on the number of neighbors~$k$ is discussed in Section~\ref{sec::clicker_exp}.

\subsection{SAINT-Q: transformer-based clicker model}
\label{sec::saint_q}
\begin{figure}[!h]
    \centering    \includegraphics[width=0.3\textwidth]{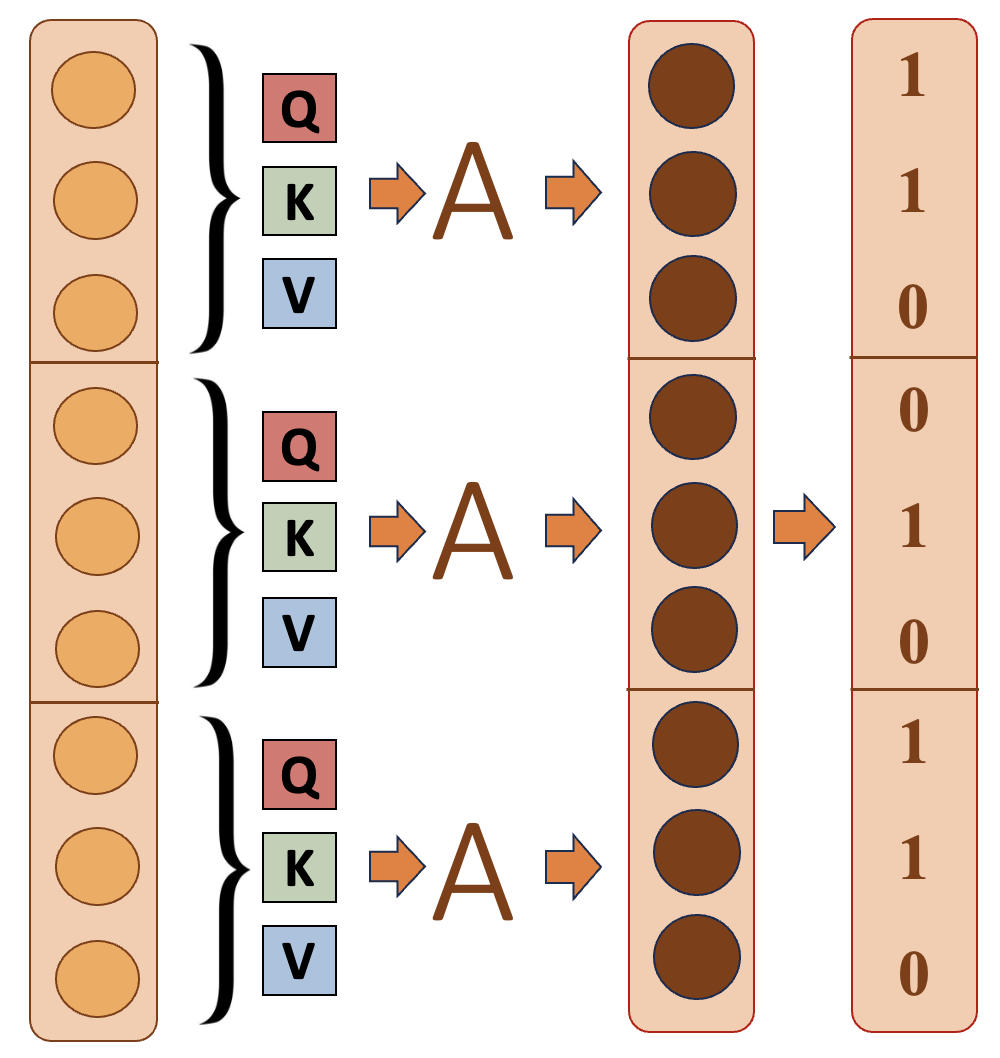}
    \caption{The illustration for the proposed Chunked Intersample Attention transformation. Tuples $(Q, K, V)$ are generated separately for different search results but with the same linear layers.
    Therefore, they are labeled with the same colors. 
    Then, these tuples are transformed via the attention mechanism and stacked for next layers processing. The output of the model is shown on the right, where $1/0$ corresponds to the predicted click.}
    \label{fig::saint_scheme}
\end{figure}

The next base model that we consider for introducing the context within the search results is SAINT~~\cite{somepalli2021saint}.
This transformer-based model for tabular data considers dependencies between features and samples simultaneously.
SAINT balances computational efficiency and performance and avoids intensive computations necessary by alternatives~\cite{padhi2021tabular,gorishniy2023tabr}.
The model supports continuous and categorical features, which is important for CTR prediction.
However, the straightforward using SAINT to solve the CTR prediction problem is impossible since it captures the dependencies between items corresponding to different queries.
Therefore, we modify the SAINT architecture to track the dependencies \emph{only within the results for a particular query}.
The resulting \emph{Chunked Intersample Attention} transformation is described formally in Algorithm~\ref{alg::inter_attention} and visually in Figure~\ref{fig::saint_scheme}.
The main feature of Chunked Intersample Attention is that the model automatically applies the attention mechanism only to the items' embeddings corresponding to the same queries.
This feature makes it possible to compose large batches with items for different queries and exploit the parallel training regime for transformer-like models~\cite{gusak2022survey}.  

\begin{algorithm}
\caption{Chunked Intersample Attention transformation. 
}

\begin{algorithmic}[1]
\State \textbf{Input}: Tensor $x$ of the size $b \times n \times d$, where $n$ is the number of features, $d$ is the head dimension, $b$ is the batch size such that $b$ is divisible by $N$.
\State Initialize output tensor $y$ of the size $b \times n \times d$
\For{$i = 0,..,b/N - 1$}
    \State $u \gets x_{[i \cdot N : (i+1) \cdot N]}$ \Comment{get chunk w.r.t. the first axis}
    \State $u \gets$ reshape($u$, $(N, n \cdot d)$) 
    \State $v \gets$ Attention($u$) 
    \Comment{Attention from~\cite{attention}}
    \State $y_{[i \cdot N : (i+1) \cdot N]} \gets$ reshape($v$, $(N, n, d)$) 
\EndFor
\State \Return $y$
\end{algorithmic}
\label{alg::inter_attention}
\end{algorithm}

\subsection{Reranking model}
\label{sec::reranker}
\begin{figure}[!h]
    \centering
    \includegraphics[width=0.4\linewidth]{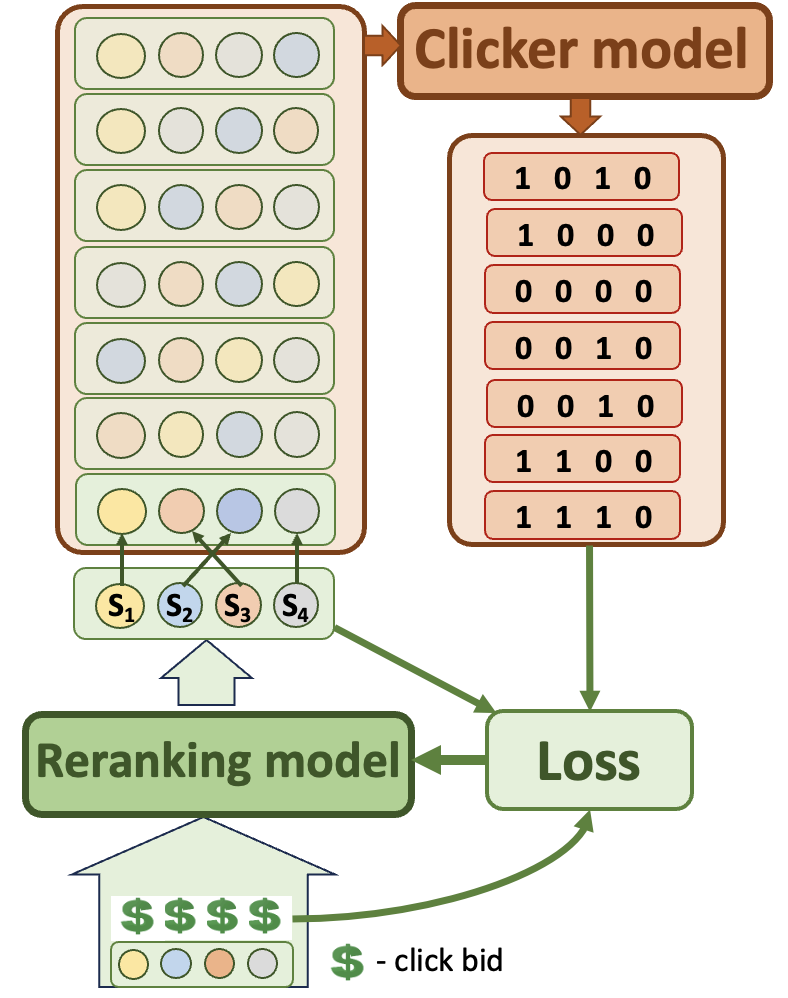}
    \caption{The reranking model training mode. Input data (the initial search query results) is passed to the reranking model, which generates scores for reranking. 
    Based on these scores, a new re-ranking is generated. 
    This ranking and all possible transpositions are fed to the clicker model, which predicts clicks on the received results. 
    These clicks, the costs per click, and the scores from the reranking model are used in the loss for training the reranking model.}
\label{fig::reranker_scheme_train}
\end{figure}

This section considers the reranking model used in our \textsf{RARe} framework.
Since the reranking model works in the online environment and processes many search results, it must generate re-ranking quickly while maintaining high performance.
Therefore, we select the MLP architecture that shows promising results if the clicker model is sufficiently accurate~\cite{2021context}.
To train the reranking model, we follow the pipeline from~\cite{2021context}, where the pre-trained clicker model with frozen parameters is used to predict user clicks on the reordered search results.
These clicks for the generated set of trial permutations~$\pi'$ are used to evaluate the loss~(\ref{eq::loss}).
The illustration of the training process is shown in Figure~\ref{fig::reranker_scheme_train}.
This scheme shows scores $s_j$, and the generated trial permutations based on single transpositions of the order induced by the scores $s_j$.
These permutations are input to the clicker model that predicts users' clicks.
Finally, these scores, predicted clicks and items' revenues are used to compute the loss and update the parameters of the reranking model. 

Note that the considered loss function could decrease the user experience from the reordered search results.
To balance commercial objectives and user experience,
authors~\cite{2020relevance} propose to adjust the items' revenue $r_i$ with the following transformation:
\begin{equation}
    r^{reg}_i = r^{organic}+\alpha \cdot r_i,
    \label{eq::r_reg}
\end{equation}
where $r^{organic}$ is the auxiliary same reward for all considered items and $\alpha \in [0, 1]$ is the tunable hyperparameter.
Although transformation~(\ref{eq::r_reg}) is initially used to balance advertising revenue and purchase volume, we transfer it to the search of the trade-off between revenue and relevance after the search results reranking.

If $\alpha= 1$ and $r^{organic}=0$, we obtain a reranking system focused solely on revenue maximization. 
If $\alpha= 0$, all items have the same expected revenue $r^{organic}$; then the system maximizes user experience measured by the total number of clicks in the top search results. 
If $\alpha$ increases with fixed $r^{organic}$, we decrease the weight of user satisfaction while enhancing the significance of revenue in the resulting ordering of search results.

\begin{figure}[!h]
    \centering
\includegraphics[width=0.3\linewidth]{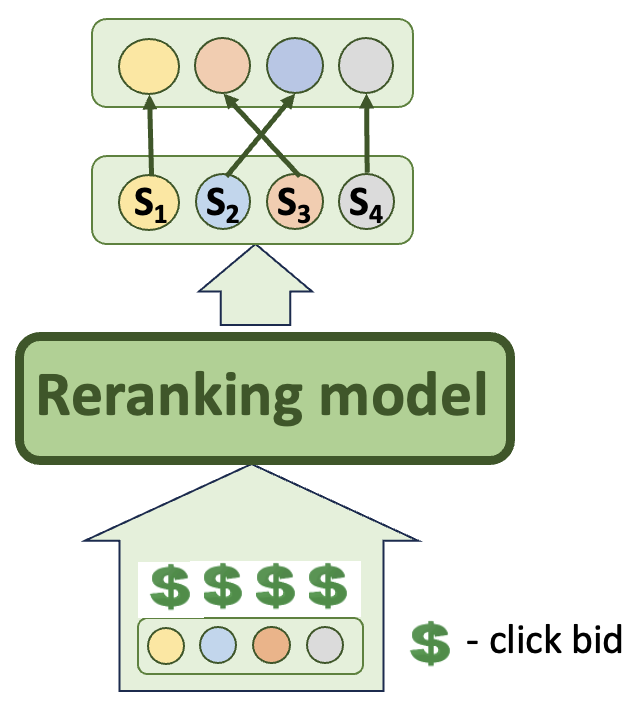}
    \caption{The scheme of the reranking model in the inference mode. 
The reranking model takes the items' features and revenues from the initial search results and reorders them to maximize the expected revenue while preserving relevance.}
    \label{fig::reranker_scheme_inf}
\end{figure}

At the same time, in the inference mode, the \textsf{RARe} framework uses only the reranking model to generate search results permutation and does not exploit any clicker model.
It inputs the items' features and revenues and produces scores $s_j$ used for reranking input items.
Therefore, the inference computational complexity of the reranking model coincides with the inference in the MLP model, which is quite low.
The illustration of the \textsf{RARe} inference mode is shown in Figure~\ref{fig::reranker_scheme_inf}.

Thus, the \textsf{RARe} framework provides a flexible mechanism for adjusting the platform's priorities, enabling us to calibrate the system according to specific business requirements while maintaining optimal performance metrics.

\section{Experimental setup}

\subsection{Evaluation metrics}
\label{sec::metrics}
Since the \textsf{RARe} framework consists of clicker and reranking models, we have to define the metrics for the evaluation performance of these models.

\paragraph{Clicker model.}
The click prediction problem can be reduced to the binary classification problem.
Therefore, we use two standard metrics to evaluate clicker models' performance: AUC and GAUC. 
The standard AUC metric measures the quality predictions over all available items and ignores the per-query performance of the clicker models.
To address this limitation, we also include GAUC~\cite{zhu2021open, zheng2022implicit} in the metric list.
GAUC averages AUC scores over separate search results:
\begin{equation}
\mathrm{GAUC} = \frac{1}{|Q|} \sum\limits_{q\in Q} AUC_q,
\end{equation}
where $AUC_q$ represents the AUC value for the items corresponding to the query $q$.

\paragraph{Reranking model.} 
To evaluate the overall performance of our framework, we show how the revenue and relevance of the search results change after their re-ordering. 
To show the increase in revenue, we use the Delta Revenue $\overline{\calR}$ metric, which estimates the gain from the perturbed order generated by the reranking model:
\begin{equation}
    \overline{\calR} = \frac{1}{|Q|} \sum\limits_{q \in Q} \frac{\widehat{\calR}_q}{\calR_q},
    \label{eq::rel_revenue}
\end{equation}
where $\calR_q$ and $\widehat{\calR}_q$ are the revenues given by original search results $I_q$ and re-ordered search results $\widehat{I}_q$, respectively.
To measure the change in relevance, we consider Difference $\calD$~(\ref{eq::kl_difference}) introduced in Section~\ref{sec::problem_statement} and Normalized Discounted Cumulative Gain (NDCG)~\cite{wang2013theoretical} between the original order of search results and the perturbed ones generated by the reranking model. 

\subsection{Baseline clicker models}

We compare our contextual clicker models with three pointwise baselines that ignore the context of the items within the search results.

\paragraph{\textbf{CTRV}.}
The CTRV model predicts the CTR based on the product of click probabilities estimated for every item individually and the decay factor, which decreases monotonically with increasing the ad's position $j$ in the search results.
This approach aligns with the model from~\cite{aggarwal2008sponsored}, where an auxiliary decay factor is used to reduce the likelihood of user interaction with items from the bottom of search results.
We use decay factor $P^{j-1}, \quad P < 1$ .
This factor makes the pointwise method more sensible to the item's position, i.e., penalizing the items from the bottom since users are less likely to click them.

\paragraph{\textbf{GBDT.}} 
Another baseline clicker model is based on the Gradient Boosting Decision Trees algorithm.
GBDT technique is commonly used for tasks involving tabular data~\cite{somepalli2021saint}, particularly for CTR prediction task~\cite{yang2022click}.
The models proposed in this study rely heavily on categorical features.
Therefore, we select the CatBoost~\cite{prokhorenkova2018catboost} implementation of the GBDT since it processes categorical data very efficiently.
At the same time, the default GBDT algorithm is not context-aware and does not track the neighbors of items for click prediction. 


\paragraph{\textbf{SAINT-S}.}
The third baseline clicker model is SAINT-S~\cite{somepalli2021saint}.
This model is the non-contextual version of the transformer model for tabular data with the standard self-attention mechanism. 


\subsection{RARe Dataset}
\label{sec::dataset}

Existing reranking datasets have significant limitations for our task. 
While the Avito Dataset~\cite{avito} provides search results with user interactions, it only contains five items per query and lacks comprehensive click-through data. 
Other benchmark datasets like Amazon~\cite{he2016ups}, LETOR~\cite{liu2007letor}, and Yahoo~\cite{chapelle2011yahoo} are inappropriate since they contain only static relevance assessments rather than real behavioral data.

We introduce the \textbf{Ra}ising \textbf{Re}venue \textbf{D}ataset(\textsf{RAReD}) containing search results with both relevance metrics and revenue data (logged clicks, click-through rates, relevance scores, ad bid prices). 
The key features of this dataset are the following:

\begin{itemize}
    \item \emph{Scale}: 68,380 search results (2,051,400 total items)
    \item \emph{Items per query}: 30 items per search result
    \item \emph{Click rate}: about 3 clicked items per search result ($CTR\approx 10\%$)
    \item \emph{Ranking}: Items in the search results are ordered by the production algorithm considering position, features, and item interactions
    \item \emph{Structure}: Items grouped by search queries
\end{itemize}

For further \textsf{RAReD} description see Appendix~\ref{app::dataset}.

\section{Clicker models experiments}
\label{sec::clicker_exp}
This section presents the comparison results for the standalone clicker models.
We use AUC and GAUC quality metrics and \textbf{RARe} dataset described in Section~\ref{sec::dataset}.
Table~\ref{tab::model_comparison} shows that the proposed models GBTD-C and SAINT-Q provide higher GAUC scores than their non-contextual base versions.
Thus, including the context within the items in the clicker models improves the click prediction performance. 
At the same time, we observe that the SAINT-Q model provides the highest GAUC score, although its inference time is larger than the inference time of the GBDT-based models.
The GBTD-C model uses $k=5$ neighbors to expand the feature space of the input dataset.
An empirical analysis of how the number of neighbors affects the model performance is presented in the next paragraph. 

\begin{table}[!h]
\centering
\caption{Performance of the proposed clicker models. 
Inference time represents the mean processing time for a search result of a size 30.
We highlight the largest scores among the transformer models with bold and underline the largest scores among GBDT-based models.}
\begin{tabular}{ccccc} 
  \toprule
  Model & AUC & GAUC & Inference time, ms \\
  \midrule
  GBDT & $\underline{0.6834 \pm 5 \cdot 10^{-3}}$ & $0.6376 \pm 5 \cdot 10^{-3}$ & $1\cdot 10^{-3}$   \\
  GBDT-C (our) & $0.6825 \pm 5 \cdot 10^{-3}$ & $\underline{0.6420 \pm 5 \cdot 10^{-3}}$ & $5 \cdot 10^{-3}$  \\
  SAINT-Q (our) & $\mathbf{0.7322 \pm 4 \cdot 10^{-4}}$ & $\mathbf{0.7549 \pm 2 \cdot 10^{-7}}$ &  $3.9$ \\ 
  SAINT-S & $0.7278 \pm 6 \cdot 10^{-4}$ & $0.7545 \pm 2 \cdot 10^{-4}$ & $6.6$ \\ 
  \bottomrule
\end{tabular}
\label{tab::model_comparison}
\end{table}


\paragraph{Number of neighbors in the GBDT-C model.}
This section demonstrates how the number of neighbors $k$ used in the data pre-processing for the GBDT-C model affects the considered AUC and GAUC metrics and inference time.
Table~\ref{tab::neighbours_comparison} shows that $k=5$ provides the highest GAUC score.
Since this metric is primary in the click prediction task, we select $k=5$ to ensure the best model for clicks simulation.
In addition, we report the inference time dependence on the number of neighbors and observe that the inference time for $k=5$ is comparable to the baseline GBDT model, making it feasible for practical applications.

\begin{table}[!h]
\centering
\caption{Comparison of different numbers of neighbors used in the GBDT-C model.  
The largest AUC and GAUC are highlighted in bold, and the second largest are underlined.}
\begin{tabular}{cccc} 
  \toprule
  \# neighbours, $k$ & AUC & GAUC & Inference time, $\mu s$ \\ 
  \midrule
  $1$ & $\mathbf{0.6834}$ & $0.6376$ & $1.2$ \\
  $3$ & $\underline{0.6831}$ & $\underline{0.6390}$ & $3.8$ \\
  $5$ & $0.6825$ & $\mathbf{0.6420}$ & $5.2$ \\
  $7$ & $0.6809$ & $0.6263$ & $5.8$ \\
  \bottomrule
\end{tabular}
\label{tab::neighbours_comparison}
\end{table}



\paragraph{Batch structure in the SAINT-Q model.}
The na\"ive implementation of the SAINT-Q processes each search result~$I_q$ independently and requires batch sizes equal to the search result length~$N$.
Our modified implementation, described in Section~\ref{sec::saint_q}, introduces \emph{Chunked Intersample Attention} (Algorithm~\ref{alg::inter_attention}), which restricts the intersample attention mechanism to operate only within items from the same query. 
While this reduces cross-query computations, it introduces chunking overhead for splitting/merging batches and separately computing each query in a batch.
Table~\ref{tab::saint_batch} demonstrates the trade-off between these effects. 
For small batches ($b<16$), the na\"ive approach outperforms our method, i.e., the chunking overhead exceeds the attention optimization benefits. 
However, processing more than $16$ queries ($b\geq16$) in parallel utilizes GPU efficiently and amortizes the chunking costs. 
Therefore, the inference time for $b=16$ matches the na\"ive baseline and larger $b$s reduce latency through better parallelism.

\begin{table}[!h]
\centering
\caption{Comparison of the SAINT-Q inference time on the test set. Every batch consists of $N\cdot b$ items. 
In our experiments $N=30$. 
Inference time for single query search results is shown in milliseconds. 
}
\begin{tabular}{ccccccc} 
  \toprule
  Number of $I_q$ in a batch, $b$ & 1 & 5 & 10 & 15 & 16 & 20 \\
  \midrule
  Inference time, ms & \textbf{8} & 16 & 10 & 9 & \textbf{8} & 7 \\
  \bottomrule
\end{tabular}
\label{tab::saint_batch}
\end{table}


\section{Reranking model experiments}

\begin{figure}[!h]
    \centering
    \begin{subfigure}[b]{0.5\textwidth}
        \centering
        \includegraphics[width=\linewidth]{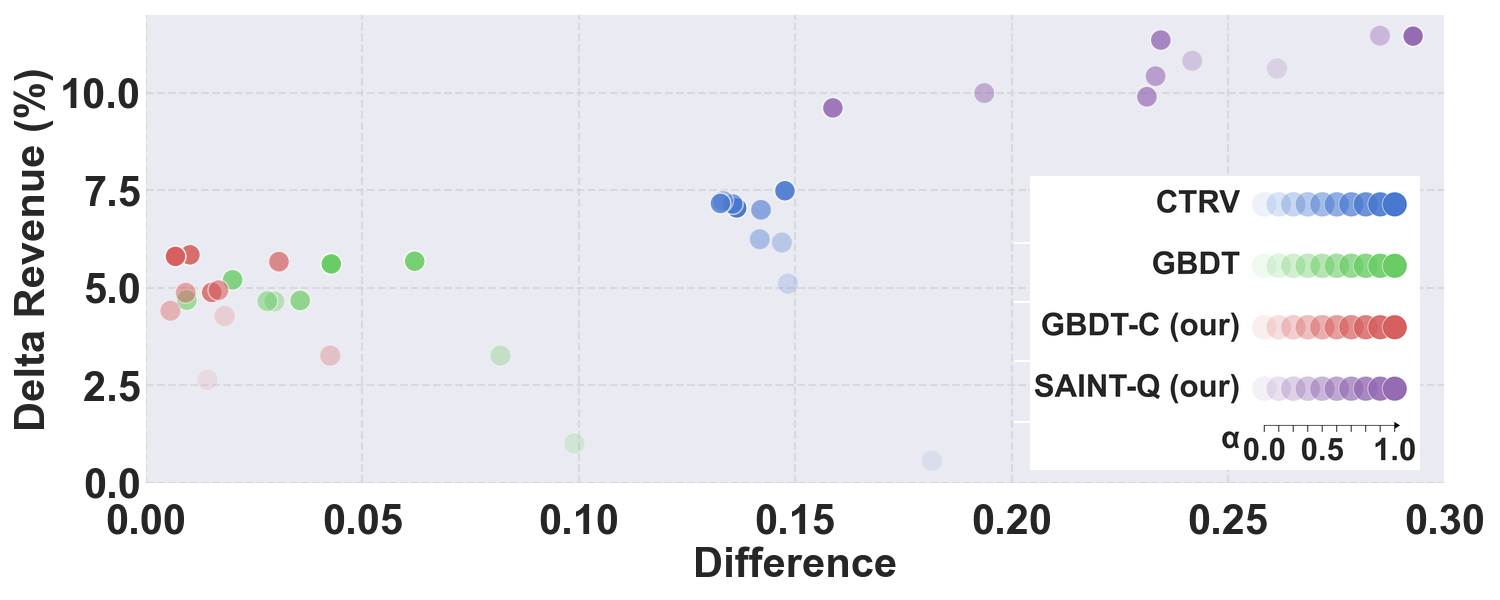}
        \caption{}
        \label{fig::res_diff}
    \end{subfigure}
    \\
    \begin{subfigure}[b]{0.5\textwidth}
        \centering
        \includegraphics[width=\linewidth]{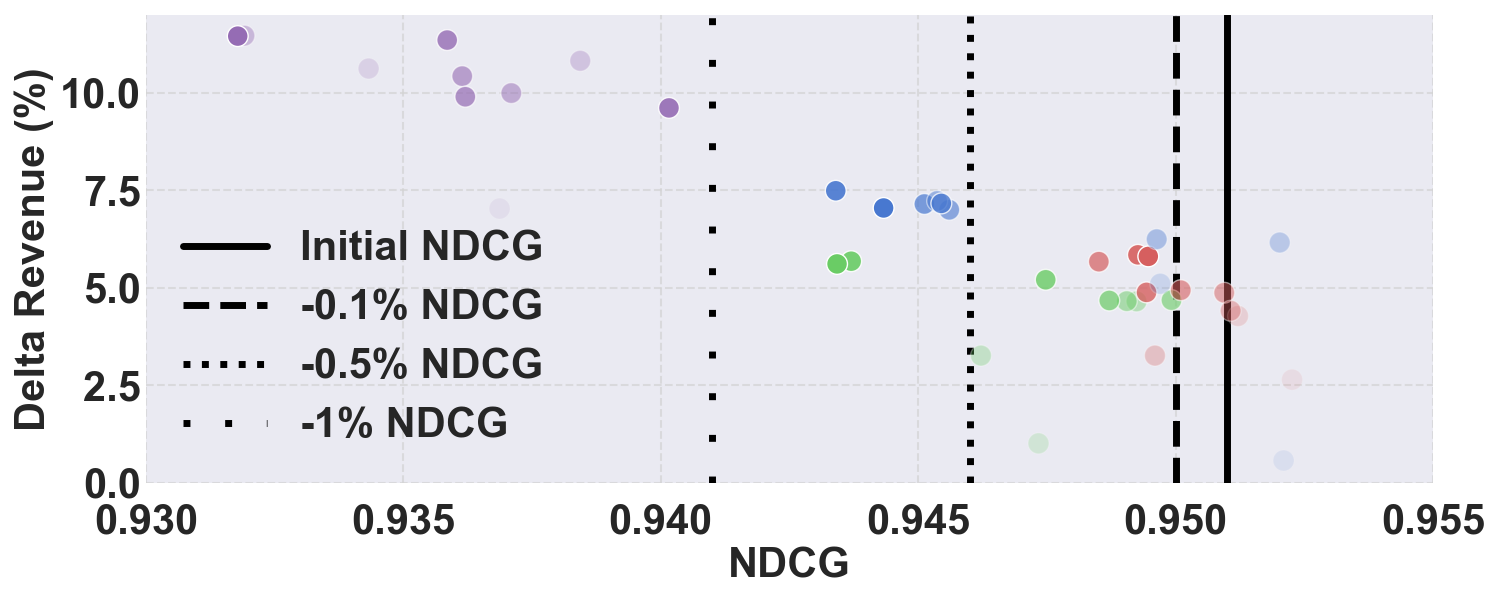}
        \caption{}
        \label{fig::res_ndcg}
    \end{subfigure}
    \caption{The Delta Revenue metric $\bar{\calR}$ and two relevance metrics for a range of hyperparameters $\alpha$ from~(\ref{eq::r_reg}).
    The larger the $\alpha$ is, the more solid the color is.
    (a) The Difference metric is used to estimate the relevance drop: the smaller the Difference, the less the order of the items is changed.
    (b) The NDCG metric estimates the relevance change compared to the original search results.
    The vertical lines show the NDCG metric at the original ranking, as well as with losses of 0.1\%, 0.5\%, and 1\%.}
\label{fig::results}
\end{figure}

This section presents the performance of the reranking models trained with considered clicker models on the test part of \textbf{RARe} dataset.
The performance of reranking models is evaluated through Delta Revenue metric~(\ref{eq::rel_revenue}) and changing the relevance of the re-ordered search results.
The latter one is measured by the Difference~(\ref{eq::kl_difference}) and NDCG metrics.
Figure~\ref{fig::results} demonstrates how the revenue adjustment transformation~(\ref{eq::r_reg}) affects the trade-off between Delta Revenue and the change in relevance.
The hyperparameter $\alpha \in [0, 1]$ controls this trade-off.
However, we observe in our experiments that the dependence of the considered metrics on the $\alpha$ is non-monotonic.

From Figure~\ref{fig::results} follows that the reranking model trained with the proposed SAINT-Q clicker model provides the largest increase in revenue (more than $10\%$) and the most significant drop in relevance (approximately $-1.5\%$ in NDCG) simultaneously.
At the same time, the ranking model based on our clicker model GBDT-C gives a smaller gain in revenue (approximately $+5.5\%$) while preserving the relevance of the search results (approximately $-0.1\%$ in NDCG).
Note that GBDT-C leads to a smaller relevance drop than GBDT while providing the same level of revenue gain.
This observation confirms the effect of our context-aware modification.
Thus, the summary of the results is the following.
First, our modification of the transformer-based model SAINT-Q provides the most significant revenue gain for the reranking model.
Second, the proposed clicker model GBDT-C preserves relevance much better than competitors while providing the same revenue gain as the non-contextual GBDT clicker. 
These results confirm that the \textsf{RARe} framework based on context-aware clicker models can either increase revenue much better than non-contextual clicker models or drop the relevance smaller with the same revenue gain.
Note that we have excluded the baseline SAINT model from such comparison since its performance as a clicker model on the \textbf{RARe} dataset is poor.

In addition to the visual comparison of the models presented in Figure~\ref{fig::results}, we provide a quantitative comparison in Table~\ref{tab::clicker_corr}.
We compute the Pearson correlation between Delta Revenue scores $\bar{\calR}$ and changing relevance metrics (the Difference $\calD$ and NDCG).
The computed correlations indicate that the reranking model based on the proposed SAINT-Q clicker model induces an increase in revenue and a drop in relevance measured by the Difference metric simultaneously.
Other models demonstrate the negative correlations between Delta Revenue and the Difference metric, indicating that the more relevance drops, the smaller the revenue gain.
The possible explanation of such an effect is that the original order of search results could be generated by a model whose primary target was relevance and the secondary target was revenue.
Therefore, significant perturbation of the order leads to a decrease in revenue and relevance simultaneously.

At the same time, the reranking model based on the GBDT-C clicker model provides a positive correlation between the Delta Revenue and NDCG. 
This correlation indicates that the smallest drop in relevance (the best NDCG score is one) corresponds to the most significant gain in revenue.
In contrast, other clicker models lead to reranking models, such that a minor decrease in NDCG leads to only a minor increase in revenue.
For these models, we expect that the larger the gain in revenue is, the more significant the drop in relevance.

\begin{table}[!h]
\centering
\caption{Pearson correlation between the Delta Revenue and relevance change metrics.}
\begin{tabular}{ccc}
\toprule
Clicker model & $\overline{\calR}$ vs. $\calD$ & $\overline{\calR}$ vs. NDCG \\
\midrule
CTRV & $-0.57$ & $-0.83$ \\
GBDT & $-0.72$ & $-0.21$ \\
GBDT-C & $-0.51$ & $0.47$ \\
SAINT-Q & $0.84$ & $-0.75$ \\
\bottomrule
\end{tabular}
\label{tab::clicker_corr}
\end{table}

\section{Conclusion}

This paper presents an efficient \textsf{RARe} framework for raising revenue by reranking search results, preserving their relevance.
It increases the revenue by 4–12\% while only a slight drop in relevance happens.
In addition, \textsf{RARe} can smoothly adjust the effect of payment increase to relevance decrease through the regularized revenue transformation.
The main features of the proposed framework are context-aware clicker models (GBDT-C and SAINT-Q), which were designed based on the existing models for tabular data.
Adapting such models to the considered revenue maximization task is the primary technical challenge we addressed in this study.
The proposed framework is tested with the relevant baselines on the industrial dataset \textbf{RAReD} that we will share with the community.
Our experiments demonstrate a significant expected increase in revenue compared to the revenues generated with the original search results.

\bibliographystyle{unsrt}
\bibliography{sample-base} 

\appendix

\section{Dataset Description}
\label{app::dataset}
Let us give a more detailed description of our dataset. It consists of one file which contains a great number of features that give full information about an item.


\textit{Query Features:}
\begin{itemize}
    \item \textbf{qid} --- ID of the search result;
    \item \textbf{hour} --- hour of the search result query (hour of request), integer variable;
    \item \textbf{platform} --- item placement platform, categorical variable (indicating whether it is the mobile or desktop version, site or app, operating system);
\end{itemize}

\textit{Individual Item Features:}
\begin{itemize}
    \item \textbf{logical\_category\_id} --- ID of the logical category of the item;
    \item \textbf{with\_delivery} --- indicates whether delivery of this item is available, binary variable;
    \item \textbf{item\_loc\_id} --- ID of the item's location;
    \item \textbf{item\_mcat\_id} --- ID of the item's subcategory;
    \item \textbf{pos\_fixed} --- position of the item within the search result, integer variable (ranging from 1 to 30);
    \item \textbf{price} --- price of the item, integer variable;
    \item \textbf{item\_cat\_id} --- ID of the item's category;
    \item \textbf{rel\_pred} --- predicted relevance according to production, float variable (ranging from 0.0 to 1.0);
    \item \textbf{cr\_pred} --- predicted conversion rate according to production, float variable (ranging from 0.0 to 1.0);
    \item \textbf{ctr\_pred} --- predicted CTR according to production, float variable (ranging from 0.0 to 1.0);
    \item \textbf{is\_auction\_winner} --- indicates whether the auction for the item is won, binary variable;
    \item \textbf{campaign\_type} --- type of interface where the daily budget and duration of the promotion are chosen by the seller, categorical variable;
    \item \textbf{xn} --- power of promotion, dependent on the daily budget, float variable;
    \item \textbf{visibility} --- predicted probability that an item will be seen at a fixed position within the search result, float variable (ranging from 0.0 to 1.0);
    \item \textbf{region} --- region of the item, categorical variable;
    \item \textbf{click\_bid} --- bid per click for auction winners, float variable;
    \item \textbf{category\_coincidence} --- indicates whether the item’s subcategory matches the query category, binary variable;
    \item \textbf{subcategory\_coincidence} --- indicates whether the item’s subcategory matches the query category, binary variable;
    \item \textbf{click} --- indicates whether there was a click on the item, binary variable.
\end{itemize}

\paragraph{Train/test split} 68,380 search results (2,051,400 total items) 
The dataset contains 68,380 search results; each search result contains 30 items.
Overall, the dataset contains 2,051,400 items in total. 
We split the dataset into training, validation, and test sets in a ratio of 0.65/0.2/0.15. 
The test set consists of 10,257 search results and 307,710 items. 

\section{Technical Implementation Details}

\subsection{GBDT and GBDT-C Models}
\label{sec:ap-catboost}

The model was run on the following set of features: \texttt{with\_delivery}, \texttt{platform}, \texttt{is\_auction\_winner}, \texttt{campaign\_type}, \texttt{pos\_fixed}, \texttt{Region}, \texttt{hour\_cat}, \texttt{price}, \texttt{ctr\_pred}, \texttt{xn}, and \texttt{click\_bid}. The first seven features were fed into the model as categorical, and the last four as continuous ones.

In the setup of GBDT-C, the result of data preprocessing for the algorithm is an expanded dataset.
Each item in this dataset is represented not only by its own features but also by features of its' neighboring items in the search result.
The division into categorical and continuous subsets also conceptually remained unchanged compared to the pre-processing for GBDT data. 
All missing values, appeared as a result of the fact that the first two and last two search result positions have an incomplete set of neighbours, in categorical columns have been transformed into strings ‘nan’, and the other values have been casted to string type, while continuous features have not undergone any changes.

Both of GBDT click models contained the same hyperparameters, namely ‘iterations’, ‘learning\_rate’, ‘depth’, ‘loss\_function’, ‘class\_weights’, ‘cat\_features’. The last three were set as ‘Logloss’, [1, 11] and a separate list of categorical columns’ names for each  GBDT click model respectively. 
The first three hyperparameters were optimized to maximize the value of the AUC metric by applying the Grid Search optimization technique, which operates by constructing a grid of hyperparameter values and evaluating the model performance for each point on the grid. 
These grids were specified as \{‘iterations’: [150, 200, 500], ‘learning\_rate’: [0.01, 0.03, 0.05, 0.1], ‘depth’: [4, 5, 6]\} and \{‘iterations’: [1000, 1500], ‘learning\_rate’: [0.01, 0.03, 0.05, 0.1], ‘depth’: [4, 5, 6]\} for GBDT and GBDT-C click models respectively. As a result, the best combinations of hyperparameters are presented in the Table 3:

\begin{table}[!h]
\centering
\caption{Optimal hyperparameters for \textbf{GBDT} and \textbf{GBDT-C} click models. The notation used: 'lr' stands for 'learning\_rate'}
\begin{tabular}{cccc} 
  \toprule
  \textbf{Click model} & \textbf{'iterations'} & \textbf{'lr'} & \textbf{'depth'}  \\
  \midrule
  \textbf{CB} & 200 & 0.05 & 5  \\ 
  {NCB} & 1000 & 0.01 & 4 \\
  \bottomrule
\end{tabular}
\label{tab::gbdt_hyper}
\end{table}

Experiments were carried out using six different seeds to increase the stability and reliability of the models.

\subsection{SAINT Transformer Click Model}
\label{sec:ap-saint}

During the experiments, the hyperparameters of the SAINT model were optimized to maximize GAUC on the validation subset. 
In this study, we leveraged the Optuna library for efficient hyperparameter optimization, conducting 60 trials with unique hyperparameter combinations to identify the optimal model configuration. The decision to limit the trials to 60 was based on the observed convergence of the optimization process, as the evaluation metric began to plateau after a certain number of iterations, indicating diminishing returns.

The hyperparameter space for SAINT model optimization is presented in Table~\ref{tab::saint_hyper}.

\begin{table}[!h]
\centering
\caption{Hyperparameter Space for SAINT Optimization.}
\begin{tabular}{cc} 
  \toprule
  \textbf{SAINT Hyperparameters} & \textbf{Search Space} \\ 
  \midrule
  Learning Rate & (1e-5, 1e-2) \\
  Embedding Dimension & [32, 64, 128] \\
  Number of Transformer Layers & [2, 8] \\
  Number of Heads & [2, 8] \\
  Attention Dropout & (0.1, 0.9) \\
  MLP Dropout & (0.1, 0.9) \\
  Loss & [CrossEntropy, LabelSmoothing] \\
  MLP hidden size &  [4, 8, 16, 32] \\
  \bottomrule
\end{tabular}
\label{tab::saint_hyper}
\end{table}

After the optimization, the best parameters were found (see Table~\ref{tab::saint_best_hyper}).

\begin{table}[!h]
\centering
\caption{Best Hyperparameters for SAINT Based on Grouped Rocauc Objective on the Validation Subset.}
\begin{tabular}{cc} 
  \toprule
  \textbf{Parameter} & \textbf{Best Value}  \\ 
  \midrule
  Learning Rate  & 1e-4 \\
  Embedding Dimension & 128 \\
  Number of Transformer Layers & 2 \\
  Number of Heads  & 4 \\
  Attention Dropout & 0.79 \\
  MLP Dropout & 0.77 \\
  Loss & LabelSmoothing \\
  MLP hidden size & 16 \\
  \bottomrule
\end{tabular}
\label{tab::saint_best_hyper}
\end{table}

To confirm the model stability, the best hyperparameters were chosen, and the training of model was conducted on 5 random seeds. Thus, the results presented in~\ref{tab::model_comparison} in the main part, including standard deviation.

\subsection{Reranking model}
\label{sec:ap-reranker}

To run the reranking model on all click models, dataset preprocessing begins with min-max normalization of the \texttt{price} field within each search result. 
The \texttt{click\_bid} values before variation by $\alpha$ are stored in a separate field, \texttt{click\_bid\_old}, for calculating the Revenue metric. 
Search results are saved in the search result size selected for the reranking task. For the reranking model (MLP), the following features are used: \texttt{with\_delivery}, \texttt{price}, \texttt{rel\_pred}, \texttt{ctr\_pred}, \texttt{cr\_pred}, \texttt{is\_auction\_winner}, \texttt{pos\_fixed}, \texttt{click\_bid}, \texttt{category\_coincidence}, \texttt{subcategory\_coincidence}, \texttt{text\_presence}, and \texttt{platform}.

The dimensions of each of thee layers of MLP for each of the clicker models are defined as hyperparameters with optimization of revenue metrics, and an output is fully connected layer with dimension 1, which produces a score for element permutation. The Adam algorithm is employed to optimize the MLP parameters, with an initial learning rate of 0.01. Weights are initialized using the Xavier method, and batch normalization is applied after each hidden layer to stabilize training. 

Hyperparameters for the reranking model are 'organic\_revenue' (corresponds to $r^{organic}$) and sizes of hidden layers (3 of them, as in original work  \cite{2021context}) in MLP. The hyperparameters were optimized to maximize Revenue value for each $\alpha$ within each user click model. SEED was fixed equal to 100 for equality of results (randomness is used only when setting the initial weights of the neural network).
 
The input had the features size (11), output size equal to 1. As the field \texttt{click\_bid} value is always on the order from ones to hundreds monetary units, it was selected ranging from 1 to 100. Hidden layer sizes were varied within [64,1024], [64, hidden\_size\_1] and [32, hidden\_size\_2] respectively. The number of trials was 50 for each configuration (user click model, $\alpha$). 

\begin{table}[!h]
\centering
\caption{Optimal reranking model hyperparameters. The notation used: HS stands for Hidden Size. $h$ is a threshold for click probabilities.}
\begin{tabular}{cccccc} 
  \toprule
  Click Model & $r^{organic}$ & HS 1 &HS 2 & HS 3 & h  \\ 
  \midrule
  CTRV & 10.0 & 640 & 146 & 53 & 0.18 \\ 
  GBDT & 50.0 & 315 & 122 &48 & 0.18\\ 
  GBDT-C & 15.0 & 220 & 100 &62 & 0.18\\
  SAINT-Q & 50.0 & 1024 & 512 & 128 & 0.22\\
  \bottomrule
\end{tabular}
\label{tab::reranker_compare}
\end{table}

The experimental results on metrics was performed after taking the average of 5 random seeds. 

\subsection{Computational environment}

All experiments were launched on: GPU model NVIDIA A100 with 16 cores, with memory capacity 128,00 GB, OS Debian 11 (Bullseye).

\end{document}